\documentclass{elsart}
\usepackage[intlimits]{amsmath}
\usepackage{epsfig}
\usepackage{amsfonts}

\newcommand{\beq}{\begin{equation}}
\newcommand{\eeq}{\end{equation}}
\newcommand{\beqa}{\begin{eqnarray}}
\newcommand{\eeqa}{\end{eqnarray}}
\newcommand{\bp}{\begin{pmatrix}}
\newcommand{\ep}{\end{pmatrix}}

\begin{document}

\begin{frontmatter}

\vspace{5mm}

\title{
{\scriptsize \tt IPPP/05/16~~~DCPT/05/32}\\
\vspace{5mm}
Solving coupled Dyson--Schwinger equations on a compact manifold
}
  \author[Dur]{C.~S.~Fischer},
  \author[Tue]{B.~Gr{\"u}ter} and
  \author[Gra]{R.~Alkofer}
  \address[Dur]{Institute for Particle Physics Phenomenology,
         University of Durham,
         Durham DH1 3LE, UK}
  \address[Tue]{Institut f\"ur Theoretische Physik,
           Universit\"at T\"ubingen,
	   Auf der Morgenstelle 14,
           72076 T\"ubingen, Germany}
  \address[Gra]{Institute of Physics,
          Graz University,
          Universit\"atsplatz 5,
          A-8010 Graz,
          Austria}
	   
\begin{abstract}
We present results for the gluon and ghost propagators in SU(N)
Yang--Mills theory on a four-torus at zero and nonzero temperatures
from a truncated set of Dyson-Schwinger equations. When
compared to continuum solutions at zero temperature sizeable modifications due
to the finite volume of the manifold, especially in the infrared, are found.
Effects due to non-vanishing temperatures $T$, on the other hand, are minute
for $T < 250\, \textrm{MeV}$.
\end{abstract}
\begin{keyword}
Nonlinear integral equations, Infrared behaviour,
Non-perturbative QCD, Running coupling,  Dyson-Schwinger equations  
\PACS  11.10.Wx, 12.38.Aw, 14.70.Dj\\
\end{keyword}

\end{frontmatter}


\section{Introduction}\label{sec:physicalproblem}

The infrared behaviour of QCD Green's functions is intimately related to the
issue of confinement, see {\it e.g.\/} ref.\ \cite{vonSmekal:2000pz}. The
confinement phenomenon is intrinsically non-perturbative, and thus a quite
limited number of methods is suited to investigate this and related issues.
Furthermore, the infrared singular nature of some of the QCD Green's functions
provides a further severe obstacle in the study of these quantities. In ref.\
\cite{Hauck:1998fz} a numerical method is described where the analytical
treatment of the gluon and ghost propagator Dyson--Schwinger equations (DSEs)
for momenta smaller than a  certain matching scale allowed for the numerical
solution of these coupled non-linear integral equations. At this point the
question arises whether a purely numerical method to solve these equations
alleviates their treatment. To this end one notes that on a compact manifold
the finite volume of underlying space-time acts as a natural infrared
regulator. Using a four-torus this will, on the one hand, be exploited here for
a direct solution of the coupled gluon and ghost DSEs, and, on the other hand,
be utilized to estimate effects of the finite volume. An extension to
asymmetric tori is equivalent to the choice of non-vanishing temperatures, and
thus it will also be considered in this paper.

As noted above confinement manifests itself in the long range behaviour of
Yang--Mills theory. According to the Kugo-Ojima confinement criterion
\cite{Nakanishi:qmas,Kugo:1979gm} as well as the Gribov-Zwanziger horizon
scenario, see {\it e.g.\/} ref.\  \cite{Zwanziger:2001kw}, it  is expected that
the gauge fixing degrees of  freedom and thus the Faddeev-Popov ghosts provide
the long-range correlations.  In Landau gauge this has been confirmed in the
framework of DSEs as well as exact renormalisation group equations in a series
of papers 
\cite{Zwanziger:2001kw,vonSmekal:1997is,Atkinson:1997tu,Alkofer:2000wg,Lerche:2002ep,Fischer:2002hn,Pawlowski:2003hq,Alkofer:2004it}: 
the ghost propagator is indeed  more singular in the infrared than a simple
pole. The gluon propagator, on the  other hand, vanishes in the infrared.
Lattice Monte-Carlo calculations agree with these findings, although it is
currently under debate whether the lattice gluon  propagator is finite
\cite{Bonnet:2001uh} or vanishes in the infrared  \cite{Silva:2004bv}. Here a
study of DSEs at finite  volumes can be helpful when judging the finite-volume
effects in lattice calculations. As due to the periodic boundary conditions 
used in lattice calculations the underlying space-time manifolds are
effectively four-tori, we will employ these manifolds in our study. In this
paper we will present general techniques needed  to solve the related DSEs, and
we will comment on the differences in the method and for the solutions when
compared to the treatment based on flat Euclidean space-time 
\cite{Hauck:1998fz,Atkinson:1997tu,Fischer:2003zc,Maas:2005xh}.  

The paper is organised as follows: In the next section we summarise general
properties of DSEs on a compact manifold. We discuss the boundary conditions of
the fields, present the equations for finite temperature and discuss the
renormalisation. Numerical results are presented in section
\ref{NR}:  Finite volume effects are discussed in subsection \ref{FV}, and
results for non-vanishing temperatures are given in section \ref{sec:results}.
We conclude with a summary and an outlook on further applications in section
\ref{con_sec}. The employed integral kernels are listed in appendix \ref{appA} and 
details of the numerical method are presented in appendix \ref{num_sec}.

\section{Dyson-Schwinger equations on a compact manifold}\label{dse_sec}

The DSEs are equations for the Green's functions, {\it i.e.\/} for the vacuum
expectation values of time-ordered products of quantum fields. Therefore it is
quite illustrative to consider first the general properties of quantum fields on
compact space-time manifolds. 

\subsection{Quantum fields on a compact Euclidean space-time manifold}

Quantum fields on a compact Euclidean space-time manifold are realised
by  imposing (anti-)\-periodic boundary conditions: 
\begin{align}
\Phi(x^\mu+L^\mu)&= e^{i2\pi \eta}\Phi(x^\mu) \; . \\ \intertext{Here $\Phi$ is
shorthand for all fields that appear in the Lagrangian of  the theory, and
$L^\mu$ denotes the length of the torus in the $\mu$-th direction. The overall
phase is $\eta=0$ for bosons, {\it i.e.\/} periodic boundary conditions,  
and $\eta=\frac12$  for fermions, {\it i.e.\/} antiperiodic boundary conditions. 
Furthermore,
in $d$-dimensional coordinate space $\Phi(x)$ has the following  Fourier
representation} \label{eq:Phiperiodic} \Phi(x) &= \frac{1}{L_1 \cdots  L_d}
\sum_{n_1 \dots n_d} \Phi_{n_1 \dots n_d} \textrm{e}^{i \omega_{\mu}
x_\mu} \; ,\\ \omega_{\mu}&=\frac{2 \pi}{L_\mu}(n_\mu+\eta) \; , 
\end{align}
in terms of its Fourier components $\Phi_{n_1 \dots n_d}$. The $\omega_\mu$
are  discretized momenta, and due to an obvious analogy we will call these 
Matsubara momenta or Matsubara frequencies. The special case we want to study
here is $d=4$, {\it i.e.\/} three space dimensions ($\mu=1,2,3$) and one 
`time` ($\mu=4$) dimension.

In the framework of the imaginary  time formalism quantum field theory at
non-vanishing temperatures is related to the corresponding Euclidean
path-integral with compactified  time-direction such that  $\Delta \tau
=\beta=1/T$. Therefore torus-compactified Euclidean space-time with a shorter
length for the time direction introduces an effective temperature which can be
estimated to be
\begin{equation}
  \label{eq:T} T=1/L_4 \; \textrm{with} \; L_4 \ll L \; .
\end{equation}  
Depending on the application we will either set all $L_\mu$ equal, or we choose 
$L_1=L_2=L_3=L$ and $L_4=\beta=1/T$. 

\begin{figure}
\begin{center} 
\epsfig{file=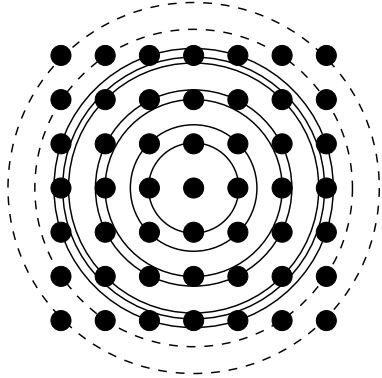,width=50mm} 
\end{center} 
\caption{Sketch
of the momentum grid dual to the four-torus. The summation over complete
hyperspheres is indicated by fully drawn circles. The hyperspheres depicted by
dashed lines are not complete if one uses Cartesian cutoffs instead of an  
$O(3)$ invariant one. Note that an $O(4)$-invariant cutoff can only be chosen
in the zero temperature case. }\label{fig:latt} 
\end{figure}

For flat Euclidean space-time the DSEs in momentum space are integral equations.
Due to the representation of the fields in Fourier space of compact space-time
we deal in this case with sums over Matsubara momenta, {\it i.e.\/} when
compactifying space-time to a torus the DSE
integrals are replaced by sums over the $n_\mu$: 
\begin{align}
\int \frac{d^4q}{(2 \pi)^4} \:(\cdots) \:\:  \longrightarrow \:\:\frac{T}{L^3}
\sum_{n_4} \sum_{n_1,n_2,n_3} \:(\cdots) \; . 
\end{align} 
For the numerical
treatment of the equations it is convenient to rearrange this
summation such that they represent a spherical coordinate  system
\cite{Fischer:2002eq},  see fig.~\ref{fig:latt} for an illustration. Employing
the $O(3)$ symmetry of the  space directions we write  
\begin{align}
\frac{T}{L^3} \sum_{n_4} \sum_{n_1,n_2,n_3} (\cdots) \: =  \frac{T}{L^3} \sum_n
\sum_{j,m} \:(\cdots) \:\; , 
\end{align} 
with $n \equiv n_4$. Furthermore, $j$ denotes spheres with $n_i n_i
=\textrm{const}$, $i=1\dots3$, and $m$ numbers the grid points on a given sphere. This
spherical summation of Cartesian grid points is very useful device when 
introducing $O(3)$  invariant cut-offs.

On a torus the gluon as a boson clearly obeys  periodic boundary  conditions.
The ghost field occurs when representing the Faddeev--Popov determinant as a
Grassmannian functional integral.  Although the ghost fields $c$ and $\bar c$
are thus anticommuting fields they also have to obey periodic boundary
conditions. Including fermionic fields $\Psi$ in the fundamental representation 
this can be seen easily from the ghost fields'  properties 
under BRST-transformations:
\begin{equation}
s \Psi = -ig t^a c^a \Psi \;. \\
\end{equation}
Here the $t^a$ are the generators of SU(N), and $s$ is the Slavnov operator. 
Since $s$ is continuous and thus unique, it is periodic. 
To preserve the antiperiodicity (due to the fermionic nature of the matter
field) on both sides of the equation periodic boundary conditions for
the ghost field $c$ are required.

\subsection{The Dyson-Schwinger Equations}

The fully renormalized zero temperature ghost and gluon 
propagators in Landau gauge can always be written in the form 
\beqa
D_{G}(k) &=& - 
\frac {G(k^2)}{k^2} \;, \label{Gh-prop}\\
D_{\mu \nu}(k) &=& 
\left(\delta_{\mu \nu} - \frac{k_\mu k_\nu}{k^2}
\right) \frac{Z(k^2)}{k^2}, \label{gluezero} 
\eeqa
where $G(k^2)$ and $Z(k^2)$ denote the ghost and gluon dressing functions,
respectively. A combination of these functions can be used to define 
the nonperturbative running coupling \cite{vonSmekal:1997is}
\beq
\alpha(k^2) = \alpha(\mu^2) \: G^2(k^2,\mu^2) \: Z(k^2,\mu^2). \label{alpha}
\eeq
which is a renormalization group invariant, {\it i.e.} independent of the 
renormalization scale $\mu$.

For non-vanishing temperatures, in the rest frame of the heat-bath,
one has two different tensor
structures in  the gluon propagator, one transverse and one longitudinal to the
heat bath \cite{Kap93,Bel96}: 
\begin{eqnarray} 
D^{ab}_{\mu \nu}(k)&=& \frac{\delta^{ab}}{k^2} 
   \bigl(P^T_{ \mu \nu}(k) \: Z_m(k_0,|\vec k|) 
        +P^L_{ \mu \nu}(k) \: Z_0(k_0,|\vec k|)\bigr) 
\quad {\rm with}	
	\label{Dmunu} \\
 && P^T_{ i j}(k)=\delta_{ij} -\frac{k_i k_j}{\vec k^2}, 
  \quad\quad P^T_{ 0 0}=P^T_{ i 0}=P^T_{ 0 i}=0, \nonumber \\
 && P^L_{ \mu \nu}(k)=P_{\mu \nu}(k)-P^T_{ \mu \nu}(k), 
  \quad P_{\mu \nu}=\delta_{\mu \nu}-\frac{k_{\mu} k_{\nu}}{k^2}\; , 
  \label{Projectors} \\ 
  && i,j=1,2,3; \quad\quad \mu,\nu=1,2,3,4, \nonumber   
\end{eqnarray}
and $k^2=k_0^2+\vec k^2$. The ghost propagator is a Lorentz scalar and does not
acquire further structures but its dressing function will depend then seperately
on frequency and three-momentum:
\begin{equation}
\label{DG} D^{ab}_G(k)=-\frac{\delta^{ab}}{k^2}G(k_0,|\vec k|).
\end{equation}
The following linear combination of gluon dressing functions,
\begin{align} \label{eq:Zz}
 Z_T(k) &:= \frac13 Z_0(k) + \frac23 Z_m(k) , 
\end{align}
will prove to be closest to
the zero temperature  gluon dressing function.

\begin{figure}
\begin{center} 
\epsfig{file=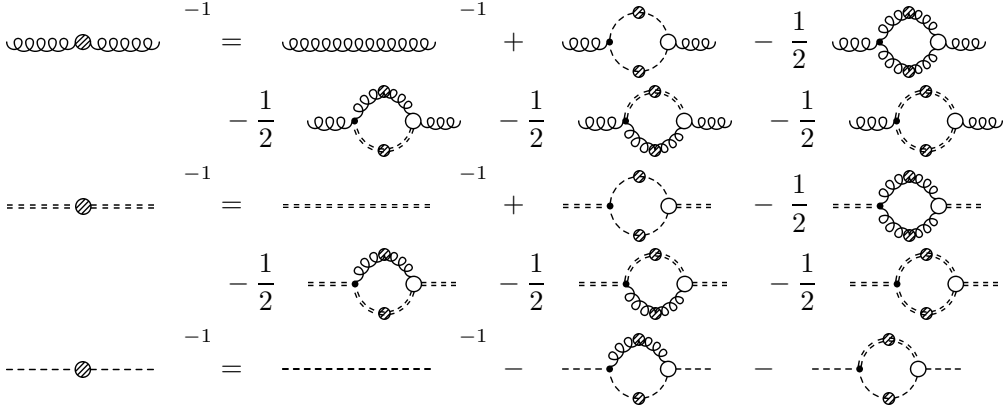,width=\textwidth} 
\end{center} 
\caption{Diagrammatic representation of the propagator DSEs in the truncation scheme 
used in this work. Wiggly lines denote heat-bath transverse gluon propagators, 
double-dashed lines are heat-bath longitudinal propagators and dashed lines 
represent ghost propagators. Blobs indicate dressed propagators and vertex functions, 
respectively.}
\label{fig:DSEs}
\end{figure}

The diagrammatic structure of the Dyson-Schwinger equations for the ghost and gluon 
propagators is shown in fig. \ref{fig:DSEs}. It is a coupled, nonlinear system of equations,
which contains dressed propagators as well as dressed vertex functions on the right hand 
side. In order to obtain a closed system of equations it is necessary to specify suitable 
approximations for these vertices. Such a truncation scheme has been developed
and successfully applied in ref.~\cite{Fischer:2002eq}. As the scheme is 
discussed in great detail in ref.~\cite{Fischer:2003zc}, we only mention briefly its main
ideas here. The key observation is ghost dominance in the infrared \cite{Alkofer:2004it}, 
{\it i.e.} the ghost loop in the gluon DSE dominates all contributions from the gluon
diagrams for small momenta. It is furthermore known, that the bare ghost-gluon vertex,
\begin{equation}
\Gamma_\mu^{gh-gl}(p,q) = i q_\mu 
\end{equation}
is an excellent approximation in both the infrared and ultraviolet momentum 
regime \cite{Cucchieri:2004sq,Schleifenbaum:2004id}. Here $q_\mu$ is the momentum of the 
outgoing ghost-leg and the colour-indices have been suppressed.  
The gluon two-loop diagrams, involving the full four-gluon vertex, can be 
neglected in these regions and it turned out to be a suitable approximations to neglect
contributions from the four-gluon vertex altogether. For the three-gluon
vertex a minimal dressing has been chosen such that the correct perturbative 
anomalous dimensions of the propagators in the ultraviolet are recovered
in the solutions. The vertex is given by
\begin{equation}
\Gamma_{\mu \nu \lambda}^{3g}(k,p,q) = \Gamma_{\mu \nu \lambda}^{3g,0}(k,p,q) H_{3g}(k,p,q)
\end{equation}
where $\Gamma_{\mu \nu \lambda}^{3g,0}(k,p,q)$ is the bare vertex. The 
dressing function $H_{3g}$ is given by 
\begin{equation} \label{3g}
H_{3g}(k,p,q)=\frac{1}{Z_1} \frac{G(q)^{(-2-6\delta)}}{Z(q)^{(1+3 \delta)}} 
\frac{G(p)^{(-2-6 \delta)}}{Z(p)^{(1+3 \delta)}}\; ,
\end{equation} 
with the momenta $q$ and $p$ running inside the loop. Here the anomalous dimension 
of the ghost propagator, $\delta=-9/44$, together with the vertex renormalization 
constant $Z_1$ ensure the correct ultraviolet running of the vertex with momenta and 
renormalization scale. Furthermore this choice leads to cutoff independent ghost and 
gluon dressing functions in the continuum. This truncation scheme has been generalised 
to non-vanishing temperatures in ref.~\cite{Maas:2004se,Gruter:2004bb,Maas:2005hs}.

On an asymmetric torus this leads to the following set of DSEs for the ghost and 
gluon dressing function $G(k)$, $Z_m(k)$ and $Z_0(k)$:
\begin{align}
\frac{1}{G(k)}   &  = \tilde Z_3 +g^2N_C\tilde Z_1 \frac{T}{L^3}  
    \sum_{n,j,m} \frac{G(q)}{k^2 q^2 (k-q)^2} 
    \left(A_T Z_m(k-q)+A_L Z_0(k-q) \right) \label{eq:ghostlong} \; , \\ 
\frac{1}{Z_m(k)} &  = Z_3 -\frac12 g^2N_C\tilde Z_1 \frac{T}{L^3} 
    \sum_{n,m,j} \frac{G(q)G(p)}{k^2 q^2 p^2} R   
   +\frac12 g^2N_C Z_1 \frac{T}{L^3} 
   \sum_{n,m,j} \frac{H_{3g}(q,p,k)}{k^2 q^2 p^2} 
   \nonumber\\
                 &  \left(M_T Z_m(q)Z_m(p)+ M_1 Z_0(q)Z_m(p)
   + M_2 Z_0(p)Z_m(q)+M_L Z_0(q)Z_0(p) \right) , \label{eq:gluontransverse}\\
\frac{1}{Z_0(k)} &  =  Z_3 -g^2N_C\tilde Z_1 \frac{T}{L^3} \sum_{n,m,j}
    \frac{G(q)G(p)}{k^2 q^2 p^2} P  + g^2N_C Z_1 \frac{T}{L^3} \sum_{n,m,j} 
    \frac{H_{3g}(q,p,k)}{k^2 q^2 p^2} 
    \nonumber \\
                 &  \left(N_T Z_m(q)Z_m(p)+ N_1 Z_0(q)Z_m(p)
    + N_2 Z_0(p)Z_m(q)+ N_L Z_0(q)Z_0(p) \right) \label{eq:gluonlongitudinal} \; .
\end{align}
The expressions for the kernel functions 
$A_T$, $A_L$, $R$, $M_T$, $M_1$, $M_2$  $M_L$, $P$ and $N_T$, $N_1$, $N_2$, $N_L$
are given in appendix \ref{appA}, see  
eqs. \eqref{eq:ghostkernels}-\eqref{eq:gluonkernelslongitudinalend}. The corresponding 
O(4)-symmetric version of these equations in the zero temperature limit 
can be found in ref. \cite{Fischer:2002eq}.

\subsection{Renormalisation}

We employ a momentum subtraction (MOM) scheme to renormalise the DSEs on the torus.
For zero temperature, {\it i.e.} on a symmetric torus, this scheme has been 
described in \cite{Fischer:2002eq}. The DSEs for the (renormalised) dressing 
functions $D \in \{G, Z_m, Z_0\}$ at the four momentum $k^2$ can be written 
symbolically as  
\begin{align} \label{sym-DSE}
  \frac{1}{D(k^2_i)} = Z^{D}+\Pi_D(k^2_i) \; ,
\end{align}
where $Z^D$ stands for the respective renormalisation constants 
$\tilde Z_3, Z_{3\, m}, Z_{3\, 0}$ and $\Pi_D(k^2)$ denote the loop 
contributions. The momenta $k_i^2$ are lying on the hyperspheres shown in 
fig.~\ref{fig:latt}. To eliminate the momentum independent renormalisation factors 
$Z^{D}$ we subtract eqs.~(\ref{sym-DSE}) from their respective equations evaluated at a 
fixed subtraction scale $s$. This scale is conveniently chosen equal to the squared 
renormalisation point, $s=\mu^2$. We obtain:
\begin{align}
  \frac{1}{D(k^2)} = \frac{1}{D(\mu^2)}+\Pi_D(k^2)-\Pi_D(\mu^2) \; .
\end{align}
One then has to specify a renormalisation condition, {\it i.e.} a value for $D(\mu^2)$,
to complete the (re-)normalisation procedure. For convenience this condition has been 
adapted from the continuum solution, given in ref.~\cite{Fischer:2002hn}. Using 
$\alpha(\mu^2)=0.968$ the renormalisation point is given by $\mu^2=1.610\,\mathrm{GeV^2}$ 
and the normalisation condition is $Z(\mu^2)=0.847$. The corresponding value of the
ghost dressing function is then $G(\mu^2)=1/\sqrt{Z(\mu^2)}=1.087$, {\it c.f.} eq.~(\ref{alpha}).

On the asymmetric torus we exploit the fact that no new divergences can arise
at finite temperatures in a renormalisable quantum field theory \cite{Kap93,Bel96}. 
Only the finite part of the renormalisation, {\it i.e.} the condition on $D(s)$,
can be affected. If the subtraction scale $s$ is much larger than 
max\{$\Lambda_{\rm QCD},1/T$\}, where the dressing functions are not affected by 
temperature effects, it should suffice to determine only one renormalisation constant 
for all Matsubara frequencies of each respective propagator. These constants are
determined {\it e.g.} from the zeroth order term in the Matsubara sum ($n=0$) by the condition
\begin{align}
  Z^{D} = \frac{1}{D(0,s)}-\Pi_D(0,s) \; ,
\end{align}
where $D(0,s)$ can be conveniently set to the zero temperature value $D(s)$.
We checked that within a very small error we obtain the same value for the constant 
$Z^{D}$ also for all other Matsubara frequencies. The overall temperature dependence of the
renormalisation constants is several percent for temperatures in the range from 
50 MeV to 800 MeV. The renormalisation constants for the heat-bath transverse and
longitudinal gluon dressing functions differ slightly, but at most a few percent
for increasing temperatures.

In regularisation schemes with an ultraviolet cutoff one also has to deal with 
quadratic divergencies, which cannot be accounted for by the MOM-procedure
described above. In the continuum formulation of the DSEs quadratic divergencies 
are removed on the level of the DSEs by a careful procedure
\cite{Fischer:2002hn,Maas:2004se} which amounts to adding a suitable counterterm to 
the bare action of the theory. A similar method has been applied to the formulation 
on the symmetric and asymmetric torus and is described in detail in 
\cite{Fischer:2002eq,Gruter:2004bb}. We therefore refrain from repeating the details
here and just note that the integral kernels given in appendix \ref{appA}
contain kernels where the dangerous terms have already been eliminated.

The dressing functions are calculated on a momentum grid that is limited by the 
ultraviolet cut-off $\Lambda$. However, as can be seen directly in equations
\eqref{eq:ghostlong}--\eqref{eq:gluonlongitudinal}, the integral kernels and the 
dressing functions in the loops depend on the relative momentum between the 
momentum $k$ from outside the loop and the loop momentum $q$. These relative
momenta may have values between zero and two times the momentum cut-off. Thus 
one has to extrapolate the dressing functions beyond the cut-off. This
is accomplished by employing the ultraviolet asymptotic forms (\ref{gluon_uv})
and (\ref{ghost_uv}), which are given in appendix \ref{sec:startguess}. These forms are known to be 
analytic solutions of the continuum
DSEs at large momenta. For typical torus sizes one does not exactly reproduce the 
one-loop scaling at large momenta, since the momentum cutoff is still rather low
of the order of a few GeV. However, we need the extrapolation only for a
small number of momentum configurations and our procedure therefore is 
satisfactory. We also checked other prescriptions, like setting the dressing functions
to a constant value or even to zero for momenta larger than the cut-off and found
only very minor changes in the solutions of the DSEs.

\section{Results}\label{NR}

\subsection{Finite volume effects at zero temperature}\label{FV}

\begin{figure}
 \begin{center}
  \epsfig{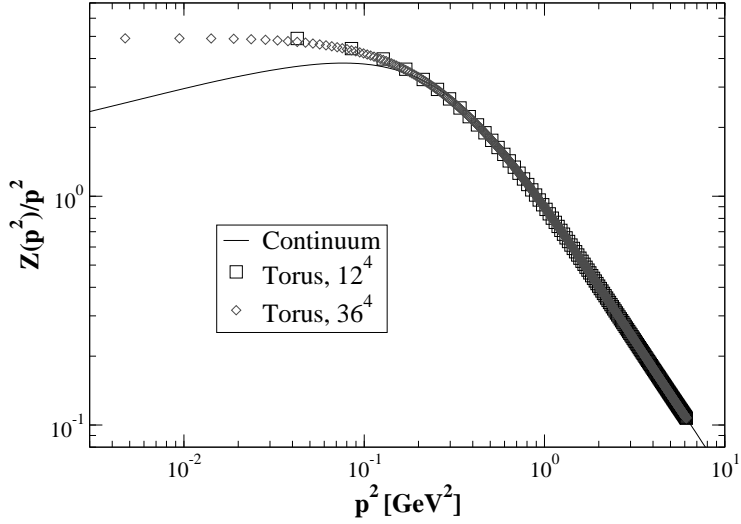}
  \end{center}
  \caption{The gluon propagator $Z(p^2)/p^2$ in the continuum and on the torus at two
  different volumes.}\label{fig:volume.D}
\end{figure}
\begin{table} 
\begin{center}
$\begin{array}{|c|c|c|c|c|c|}
\hline
\Lambda [\textrm{GeV}]    &\ 2.47 \ &\ 3.09 \ &\ 3.71\ &\ 4.33 \ &\ 4.95 \  \\ \hline
D(0) [\textrm{GeV}^{-2}]  &   4.90  &  4.55   &  4.23  &  3.97   &  3.75    \\ \hline
\end{array}$
\caption{Finite size analysis: Dependence of $D(0)$ on the cutoff $\Lambda$ at a fixed volume of 
$V = 4091 \,\mathrm{fm}^4$.}\label{fittable}
\end{center}
\end{table}

\begin{figure}
 \begin{center}
  \epsfig{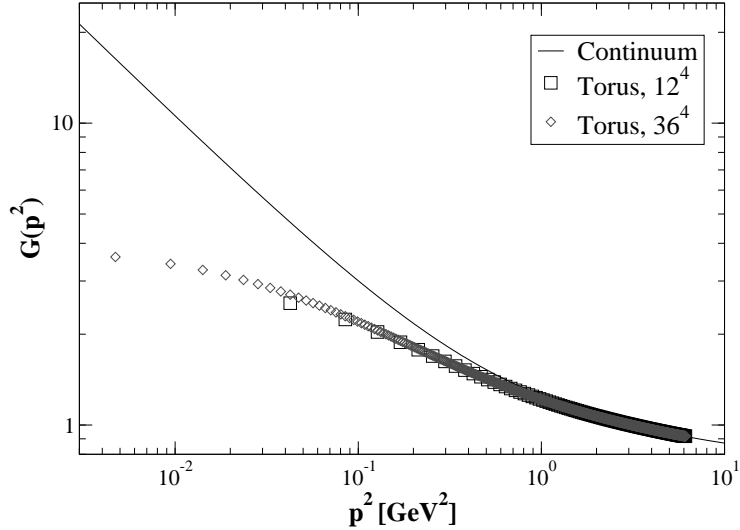}
  \end{center}
  \caption{The ghost dressing function $G(p^2)$ in the continuum and on the torus at two
  different volumes.}\label{fig:volume.G}
\end{figure}

Our first application of the numerical tools described in the previous sections is to 
determine whether one can recover the solutions of the continuum DSEs, determined in 
ref.~\cite{Fischer:2002hn}, also on a torus. First results of such an investigation 
have already been reported in refs.\ \cite{Fischer:2002hn,Fischer:2004ym}. Here we extend 
this investigation to much larger volumes seeking for signals of the continuum limit 
on the torus. To clarify the effects in the most pronounced way we display results for 
the gluon {\it propagator}, $D(p^2) = Z(p^2)/p^2$, in fig. \ref{fig:volume.D}  and the ghost 
{\it dressing function}, $G(p^2)$, in fig. \ref{fig:volume.G}. The momentum scale on 
the torus is adapted to the one from the corresponding continuum calculation 
\cite{Fischer:2002eq}. Our strategy was to fix the cutoff at a certain scale, 
$\Lambda=2.47$ GeV for the solutions presented in figures \ref{fig:volume.D} and 
\ref{fig:volume.G}, and then vary the numbers of points on the momentum lattice to 
generate different volumes of the torus. For $p_0$ in units of GeV and denoting the 
smallest momentum accessible on the (symmetric) lattice, the corresponding volume is 
determined by $V = (2 \pi/p_0 \cdot 0.1973 \, \mathrm{fm})^4 $. The two solutions displayed
in figures \ref{fig:volume.D} and \ref{fig:volume.G} have been calculated with 
$V \approx 2500 \, \mathrm{fm}^4$ and $V \approx 10^5 \, \mathrm{fm}^4$. Note that the smaller 
volume corresponds to a typical one achieved in recent lattice simulations 
\cite{Bonnet:2001uh},
whereas the larger value exceeds the current possibilies of the lattice formulation by 
far. Interestingly, our results for these two volumes are very similiar to each other, 
whereas there is a qualitative gap to the continuum results. The numerical continuum 
solution for the gluon propagator is in accordance with the infrared analytical result,
eq. (\ref{powerlaw}), and clearly shows a vanishing gluon propagator at zero momentum
with the anomalous dimension $\kappa \approx 0.595$.
The solutions on the torus, on the other hand, seem to go to a constant if extrapolated 
to zero momentum. 

\begin{table} 
\begin{center}
$\begin{array}{|c|c|c|c|c|c|}
\hline
V [\textrm{fm}^4]    &\ 105964 \ &\ 51102 \ &\ 20931\ &\ 4091 \ &\ 2461 \  \\ \hline
G(0)                 &   3.84    &  3.79    &  3.68   &  3.36   &  3.10    \\ \hline
\end{array}$
\caption{Finite volume analysis: Dependence of $G(0)$ on the volume $V$ at a fixed cutoff of 
$\Lambda = 2.47$ GeV.}\label{fittable2}
\end{center}
\end{table}

In order to analyse the zero momentum limit more precisely we fitted the infrared part 
of our solution to the form 
\beq \label{IR-fit}
\frac{Z_{ir}(p^2)}{p^2} = 
\frac{a_0}{p^2} \left(\frac{p^2}{a_1^2 + p^2}\right)^{2 a_2}.
\eeq
For momenta below $p \approx 250$ MeV we found very good agreement with our large
volume solution for the values of the parameters $a_0 = 0.244$, $a_1 = 0.233$ GeV 
and $a_2 = 0.500$. Thus, to very good precision we find that the torus solution
for the gluon propagator is constant in the infrared, and thus very different from
its continuum limit. However, there is a caveat here.
Finite size effects due to the ultraviolet momentum cutoff $\Lambda_{UV}$ could introduce 
contributions of the order $1/\Lambda_{UV}$ or $1/\Lambda_{UV}^2$ to $D(0)$ on the torus. 
Provided these terms have large coefficients they could dominate even for very small momenta
and hide a possible underlying power law with $\kappa > 0.5$. In order to test for finite 
size effects we solved the DSEs on a torus with fixed volume $V = 4091 \,\mathrm{fm}^4$
for five different cutoffs corresponding to momentum lattices of $16^4, 20^4, 24^4, 28^4$ and
$32^4$. We then fitted the expression (\ref{IR-fit}) in the infrared with $a_2$ fixed to
$0.5$. Our results for $D(0)$ are given in table \ref{fittable}. We find a considerable 
dependence of $D(0)$ on the cutoff, which can be described by the form
\beq
D(0) = d_0 + \frac{d_1}{\Lambda_{UV}}
\eeq
with the parameters $d_0=2.67$GeV$^{-2}$ and $d_1=5.57$GeV$^{-1}$. 
If we use $\Lambda_{UV}^2$ instead of 
$\Lambda_{UV}$ the quality of the fit gets much worse, indicating that the dominant 
finite size effects are of order $1/\Lambda_{UV}$. Nevertheless, the nonvanishing, 
cutoff independent constant $d_0$ indicates that the gluon propagator on a torus 
is indeed finite in the infrared.

\begin{figure}
 \begin{center}
  \epsfig{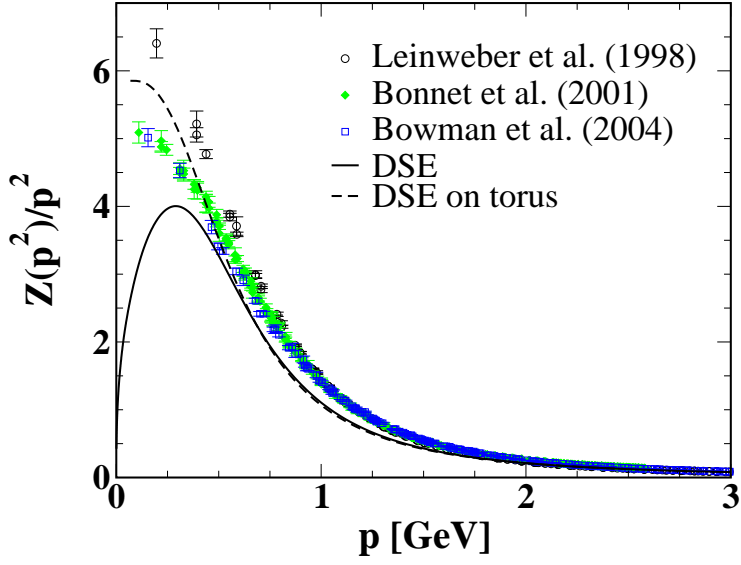}
  \end{center}
  \caption{The gluon propagator $Z(p^2)/p^2$ in the continuum and on the torus compared to the lattice results
  of refs. \cite{Bonnet:2001uh,Leinweber:1998im,Bowman:2004jm}. The most recent lattice data from 
  ref.~\cite{Sternbeck:2005tk} agree well with the data shown.}\label{volume.D.latt.eps}
\end{figure}

\begin{figure}
 \begin{center}
  \epsfig{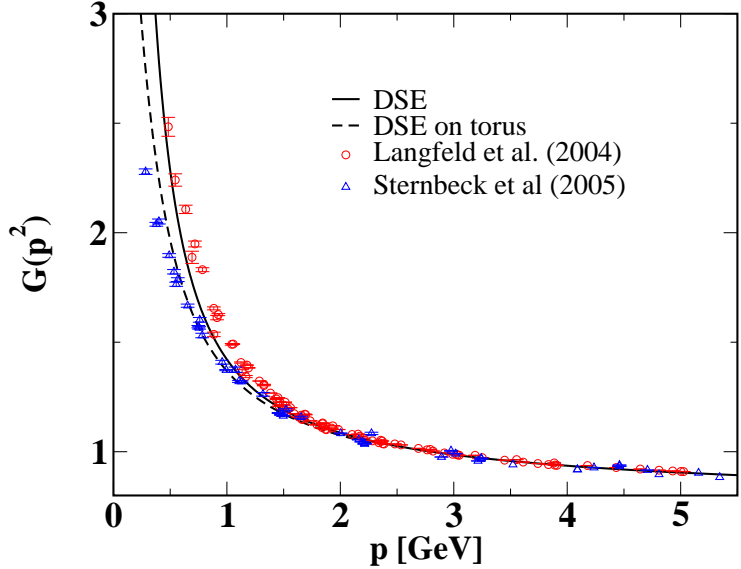}
  \end{center}
  \caption{The ghost dressing function $G(p^2)$ in the continuum and on the torus compared to the lattice results
  of refs. \cite{Sternbeck:2005tk,Gattnar:2004bf}.}\label{volume.G.latt.eps}
\end{figure}
 
Our solutions for the ghost dressing function on a torus are shown in fig. \ref{fig:volume.G}. 
Similar to the gluon we note a clear qualitative difference between the torus and the
continuum solutions. Although there is a distinct increase of $G(p^2)$ in the midmomentum region,
the function is much less steep in the infrared than the continuum solution. The volume 
dependence for a given cutoff is very small and it is not at all clear that 
$G(0) \rightarrow \infty$ for $V \rightarrow \infty$. On the contrary, the infrared part of the 
ghost dressing function for the large volume can be fitted very well with the form
\beq
G_{ir}(p^2) = \frac{e_0}{(e_1 + p^2)^{e_2}}, \label{IR-fit2}
\eeq
where  
$e_0 = 0.11$GeV$^{2e_2}$, $e_1 = 0.02 \mbox{GeV}^2$ and $e_2 = 0.32$. If the scale $e_1$ is not chosen
as a fit parameter but set to zero the fit becomes unstable. This clearly indicates that the 
dressing function is finite when extrapolated to zero momentum. 
From this fit form we can determine $G(0)$ for several volumes. Our results are given in table 
\ref{fittable2}. From these values it is not clear whether the ghost diverges in the infinite 
volume limit. If so, then this limit is approached at an extremely slow rate. A possible polynomial 
representation of the values in table \ref{fittable2} is given by $G(0) = g_0 - g_1/V^{g_2}$ with
$g_0=3.89$, $g_1=0.305$ and $g_2=0.668$, which indeed indicates a constant ghost in the
infinite volume limit. However, other forms may fit the values equally well.

In figs. \ref{volume.D.latt.eps} and \ref{volume.G.latt.eps} we compare our results for the gluon propagator
and the ghost dressing function with corresponding ones from 
recent lattice calculations \cite{Bonnet:2001uh,Leinweber:1998im,Bowman:2004jm,Sternbeck:2005tk,Gattnar:2004bf}
(see also \cite{Oliveira:2004gy} for first results on large asymmetric lattices). 
Interestingly enough, there is a striking qualitative agreement between the lattice data and our torus result
for the gluon propagator.
Both seem to go to a finite value in the infrared, in disagreement with the continuum solution.
Thus, provided the analogy between the lattice formulation and the formulation of the DSEs on the torus holds,
then the 'true', continuum limit gluon propagator may look quite different in the infrared than the one of 
contemporary lattice calculations. As for the ghost, both results, the DSE solution in the continuum and the 
one on the torus look strikingly similar on a linear plot of the momentum range assessible on the lattice. 
In turn, this means that the lattice ghost propagator could equally well be
divergent or finite at zero momentum. 

\begin{figure}
 \begin{center}
  \epsfig{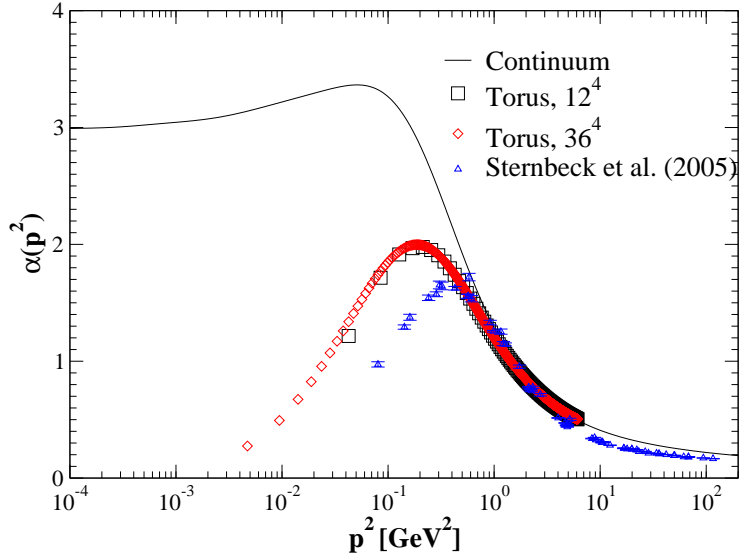}
  \end{center}
  \caption{The running coupling in the continuum and on the torus 
  compared to the lattice results of ref. \cite{Sternbeck:2005tk}.}\label{volume.alpha.eps}
\end{figure}

The running coupling (\ref{alpha}), determined from a combination of the ghost and gluon dressing functions,
is shown in fig. \ref{volume.alpha.eps}. Whereas in the continuum the coupling approaches the infrared fixed 
point (\ref{fixedpoint}) for small momenta, we find an infrared vanishing coupling on the torus. Note that the
difference between the small volume ($12^4$) and large volume ($36^4$) result on the torus is very small.
Thus one cannot extrapolate to the continuum limit by varying the torus volume. Compared to the results 
of recent lattice calculations \cite{Sternbeck:2005tk} we again find qualitative agreement between the torus solutions
and the lattice data\footnote{The quantitative difference between the two approaches at large momenta is 
due to missing two-loop contributions in the gluon-DSE.}.  

\subsection{Ghost and gluon propagators at temperatures below $T_c$}
\label{sec:results}

\begin{figure}
 \begin{center}
  \epsfig{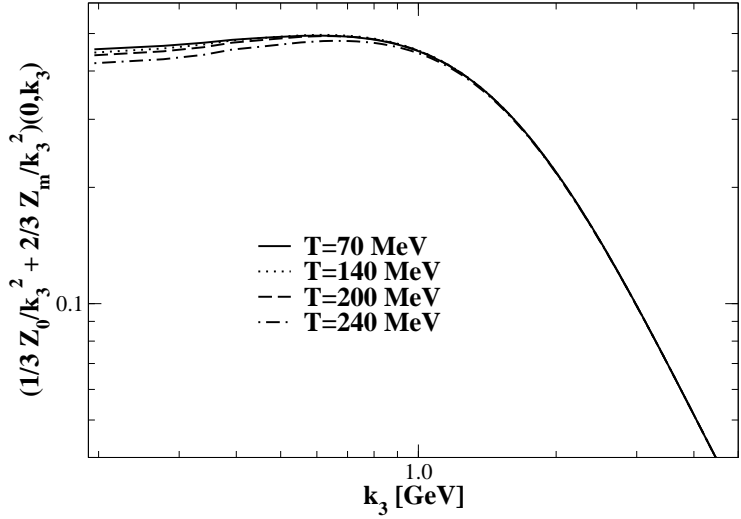}
  \end{center}
  \caption{The zeroth Matsubara component of the gluon propagator, $Z_T/k_3^2$,
  in the infrared region at different
  temperatures ($24^3$ momentum grid).}\label{fig:Z_24_70-240}
\end{figure}

\begin{figure}
\begin{center}
  \epsfig{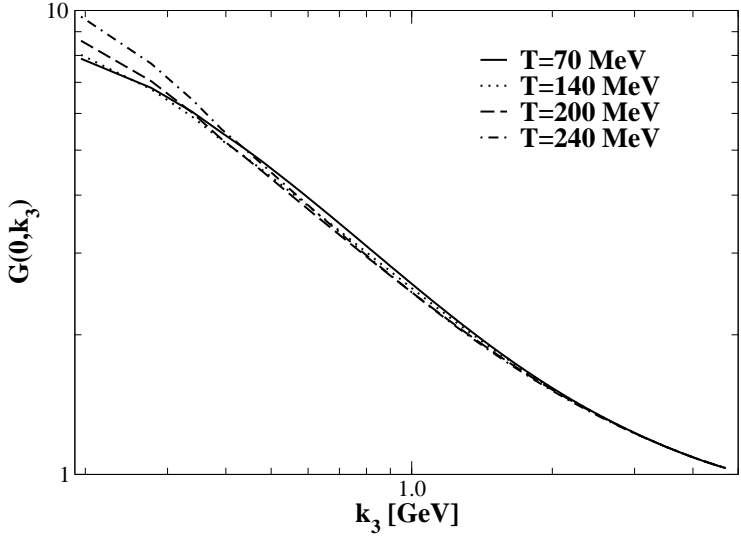}
\end{center}
 \caption{The ghost dressing function $G$ in the infrared region at different temperatures
 ($24^3$ momentum grid).}\label{fig:G_24_70-240}
\end{figure}

A convenient way to assess the temperature dependence of the gluon dressing functions 
is to evaluate the linear combination  $Z_T(k)$, eq.~(\ref{eq:Zz}), and the difference
of the dressing functions $Z_0$ and $Z_m$,
\begin{align} \label{eq:DeltaZ}
  \Delta Z (k) &= Z_0(k)  - Z_m (k). 
\end{align}
The linear combination (\ref{eq:Zz}) connects directly to the gluon dressing
function $Z(p^2)$ at zero temperature whereas $\Delta Z$ measures temperature
effects on the tensor structure  of the gluon propagator. 

The obtained results have in common that the ghost and gluon propagators
are only slightly temperature dependent for $T\le 240$MeV.  Hereby the
most interesting parts are their zeroth Matsubara modes since all other modes
turn out to be $O(4)$ invariant, {\it i.e.\/}  they depend to a very high
precision only on $p^2=\vec p^2 + p_4^2$, and therefore show no explicit
temperature effects.  

\begin{figure}
\begin{center}
  \epsfig{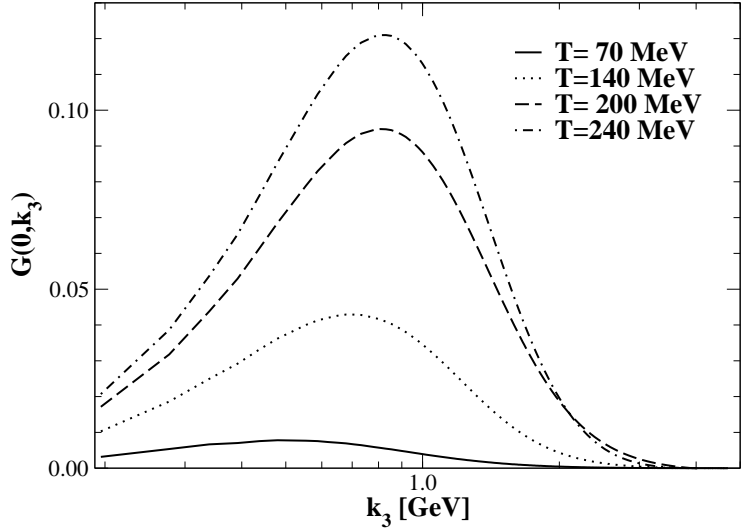}
\end{center}
 \caption{The difference $\Delta Z$ of the gluon dressing functions at different
 temperatures ($24^3$ momentum grid).} \label{fig:DZ_24_70-240}
\end{figure}

\begin{figure}
\begin{center}
  \epsfig{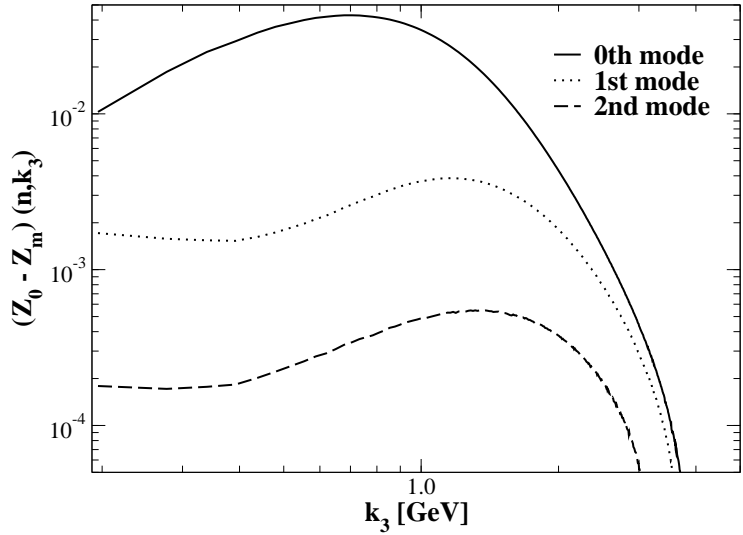}
\end{center}
 \caption{Different Matsubara modes of the difference $\Delta Z$ of the gluon 
 dressing functions at $T=140$ MeV ($24^3$ momentum
 grid).} \label{fig:ZM_140_modes}
\end{figure}

The results presented in figs.\ \ref{fig:Z_24_70-240} and \ref{fig:G_24_70-240}
have been calculated using a $24^3$ grid and an ultraviolet cutoff of 4.7 GeV.
This corresponds to a spatial volume of approximately 6 fm$^3$.  In figs.\
\ref{fig:Z_24_70-240} and \ref{fig:G_24_70-240} we show the zeroth 
Matsubara modes of the ghost dressing function $G$ and the quantity 
$Z_T/k_3^2$ at several temperatures.  As has been discussed in the
previous section only the finite-volume value of the correspondig infrared 
exponents can be extracted. The propagator function $Z_T/k_3^2$ 
shows only small variations for a large range of temperatures up to at least 240 MeV.  
The ghost dressing functions also changes only by a few percent. 
The quantity $\Delta Z$ (\ref{eq:DeltaZ}) which 
measures the change of the tensor structur of the gluon propagator is 
displayed in fig.~\ref{fig:DZ_24_70-240}. 
Whereas $\Delta Z$ remains quite small for small
temperatures, it increases drastically in the mid-momentum regime around
$k_3=700$ MeV when the temperature is raised above 200 MeV, 
thereby indicating a sizeable
change of the gluon propagator in this region. These temperature effects are
presumably most important just below the critical temperature.  
Thus the question arises whether higher Matsubara modes are important for this
quantity. As one sees from fig.~\ref{fig:ZM_140_modes} this is definitely not
the case for T=140 MeV. At larger temperatures we find even smaller contributions
from these modes.

\begin{figure}
\begin{center}
  \epsfig{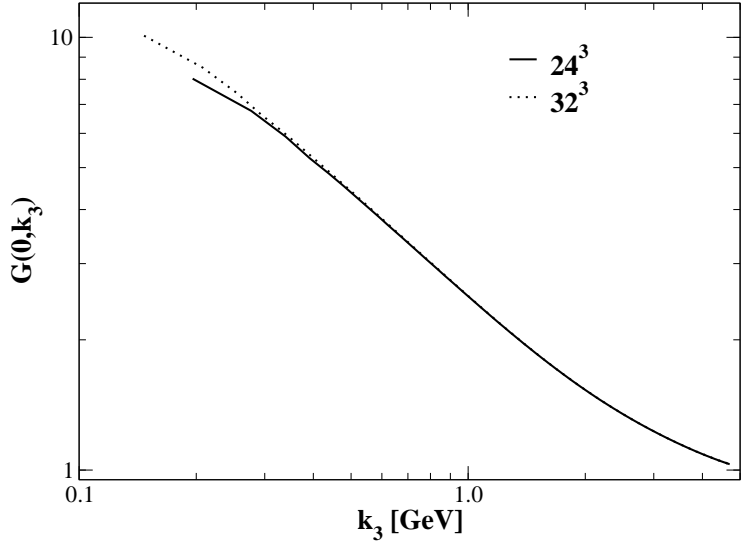}
\end{center}
 \caption{The ghost dressing function $G$ in the infrared region from 
 different grid sizes at T=140 MeV).}\label{fig:G_fs_140}
\end{figure}

\begin{figure}
 \begin{center}
  \epsfig{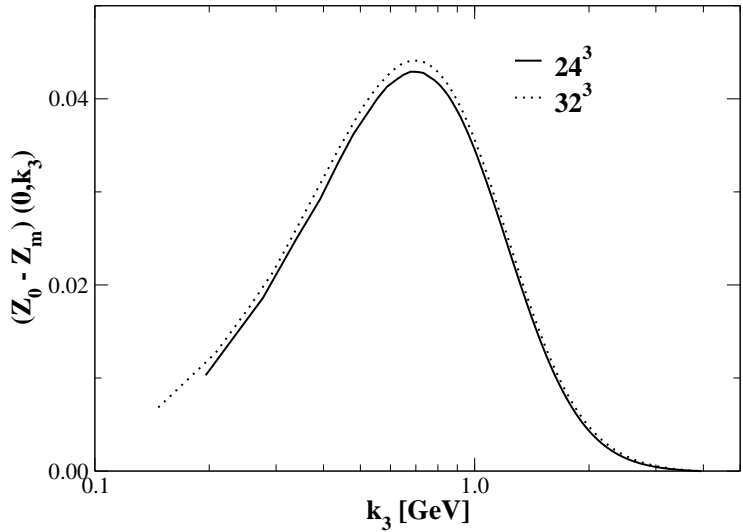}
  \end{center}
  \caption{The difference $\Delta Z$ of the gluon dressing functions 
  from different grid sizes at T=140 MeV}
  \label{fig:DZ_fs_140}
\end{figure}

At T=140 MeV we have also studied the system for considerably larger volumes. 
As one can see from figs.\ \ref{fig:G_fs_140} and \ref{fig:DZ_fs_140}  the
changes when going to larger volumes are, even in the infrared, minute. Of
course, this has to be expected from the results presented in previous
subsection. Nevertheless, as discussed this does not mean that the
infinite-volume limit of these quantities is reached.

Finally, we want to mention that with this numerical method it is
possible to find qualitatively similar solutions for the gluon and ghost 
dressing functions also for quite high temperatures up to T=800 MeV
\cite{Gruter:2004bb}. At the moment we lack a definite  interpretation of such
solutions, the most likely explanation is that they belong to a super-heating
of the low-temperature phase.

\section{Summary}\label{con_sec}

We presented solutions for the coupled set of  Dyson-Schwinger
equations for the ghost and gluon propagators of  Landau gauge Yang-Mills
theory on a four-dimensional torus. We studied the influence of the finite 
volume on the infrared behaviour of the propagators on a symmetric torus
and we employed asymmetric tori to introduce non-vanishing temperatures.

Evaluating the ghost and gluon propagators on the torus for various  volumes
and momentum cutoffs provides a surprising result: Even for extremely large
volumes, exceeding contemporary lattice volumes by  magnitudes, distinct
qualitative differences between torus and  continuum solutions are found. On
the torus, the gluon propagator as well as  the ghost dressing function are
finite in the infrared, whereas the  same quantities in the continuum limit are
infrared vanishing (gluon)  or diverging (ghost). By varying the ultraviolet
cutoff we made sure  that these effects are not artefacts of the ultraviolet 
regularization. Thus the continuum limit seems to be a
nontrivial transition from the compact to the noncompact manifold. 

We furthermore compared our results with contemporary lattice 
calculations. For the ghost dressing function all three solutions look 
similar in the momentum range accessible by lattice calculations. 
For the gluon propagator, however, we found qualitative agreement of 
the torus results with the lattice data in the infrared, whereas the
continuum result is different. Thus our results
suggest that the continuum gluon propagator may be qualitatively 
different than the one found in contemporary lattice calculations. 
The same is true for the running coupling. The analytical results from the
continuum DSE-approach give a fixed point in the infrared. Although the 
value of this fixed point is, within narrow limits, subject to the employed ansatz for the 
ghost-gluon vertex \cite{Lerche:2002ep}, its very existence can be derived 
from the structure of the ghost-DSE alone and is therefore 
independent of a truncation \cite{Watson:2001yv}. The fixed point is universal 
\cite{Alkofer:2004it} and shown to be invariant in a class of gauges that 
interpolate between Landau and Coulomb gauge \cite{Fischer:2005qe}. On a 
compact manifold, however, both, the DSE-approach and lattice Monte-Carlo 
simulations \cite{Sternbeck:2005tk,Furui:2003jr,Alles:1996ka,Boucaud:2005qf} 
find a vanishing coupling 
in the infrared in disagreement with the analytically derived continuum limit. 

A further problem studied in this work is the effect of non-vanishing
temperatures on the infrared anomalous  dimensions of the ghost and gluon
propagators. Our results indicate that the anomalous dimensions are unaffected
up to temperatures of the order of the phase transition.  Moderate-size
differences do occur in the tensor structures of the  gluon propagator
indicating a qualitative change in the gluon propagator as one approaches the
phase transition from below. Nevertheless, it remains an open question which
degrees of freedom are the most relevant for the deconfinement phase
transition.  

We presented a numerical method to solve coupled sets of  nonlinear integral 
equations on a compact manifold by iteration.  The method is suitable for
both symmetric and asymmetric  space-time tori. It can be applied to systems
with an arbitrary  number of dimensions and an arbitrary number of equations. 
We are confident that the algorithm presented here also provides a sound basis for
investigations of related issues. {\it E.g.\/} the choice of twisted boundary
conditions on the torus \cite{Tok:2005ef} might hold the key to provide a tool 
for assessing the
influence of topologically non-trivial field configurations on the infrared
behaviour of QCD Green's functions.

\section*{Acknowledgements}
We thank A.~Maas, P.~Maris, O.~Oliveira, P.~Silva and A.~G.~Williams for helpful discussions, and 
T.~Zibold for bringing the Gauss--Seidel algorithm to our 
attention. We are indepted to Andre Sternbeck for communicating 
the lattice results of ref.~\cite{Sternbeck:2005tk} to us.
Christian Fischer is grateful for the hospitality of the group 
at the University of Coimbra where part of this work was done.

This work has been supported by the DFG under contracts FI970/2-1 and GRK683
(European Graduate  School T{\"u}bingen-Basel) and by the Helmholtz
association (Virtual Theory Institute VH-VI-041).

\begin{appendix}

\section{Integral kernels for the Dyson-Schwinger Equations}\label{appA}

The two kernels of the ghost equation \eqref{eq:ghostlong} are given by:
\begin{align} \label{eq:ghostkernels}
A_T(k,q)& = 
-\frac{\vec k^2 \vec q^2 - (\vec k \cdot \vec q)^2}{(\Vec k-\Vec q)^2} 
\end{align}
\begin{align}
A_L(k,q)&=  -\frac{k^2  q^2 - (k q)^2}{( k- q)^2} +
\frac{\Vec k^2 \Vec q^2 - (\Vec k \cdot \Vec q)^2}{(\Vec k-\Vec q)^2} .
\end{align}
The kernel of the ghost loop and the four kernels of the gluon loop for the 
heat-bath transversal part of the gluon equation \eqref{eq:gluontransverse} are:
\begin{align} \label{eq:gluonkernelstransverse}
R(k,q)&=-\frac{\left(\vec q^2 \vec k^2 -(\vec k \cdot \vec q)^2 \right)}{\vec k^2} \\
M_T(k,q)&=-2\frac{\vec q^2 \vec k ^2 -(\vec k \cdot \vec q)^2}{\vec k^2 \vec q^2 \vec p^2} 
\left((\vec k \cdot \vec q)^2+\vec k^2 \vec q^2 + 
 2\vec p^2(\vec k^2 + \vec q^2 )\right) \\
M_1(k,q)&=-2 \frac
{  \left(q_0 \vec k \cdot \vec q - k_0 \vec q^2 \right)^2
\left((\vec k \cdot \vec p )^2 + \vec k^2 \vec p^2\right) }
{\vec k^2 \vec q^2  \vec p^2 q^2}\\
M_2(k,q)&=-2 \frac
{  \left(p_0 \vec k \cdot \vec p - k_0 \vec p^2 \right)^2
\left((\vec k \cdot \vec q )^2 + \vec k^2 \vec q^2\right) }
{\vec k^2 \vec q^2  \vec p^2 p^2}\\
M_L(k,q)&= -2\frac{\left(\vec q^2 \vec k^2-(\vec k \cdot \vec q)^2 \right)
\left(q^2 p^2-q_0 p_0 qp \right)^2}
{\vec k^2 \vec q^2 \vec p^2 k^2 p^2} .
\end{align}
The kernels for the heat-bath longitudinal part \eqref{eq:gluonlongitudinal} 
have the form:
\begin{align} \label{eq:gluonkernelslongitudinal}
P(k,q)&=-\frac{\left(q_0\vec k^2  -k_0\vec k \cdot \vec q \right)^2}{k^2 \vec k^2} \\
N_T(k,q)&=-2 \frac
{  \left(k_0 \vec k \cdot \vec q - q_0 \vec k^2 \right)^2
\left((\vec q \cdot \vec p )^2 + \vec q^2 \vec p^2\right) }
{\vec k^2 \vec q^2 \vec p^2 k^2}\\
N_1(k,q)&= -2\frac{ \left( \vec q^2 \vec k^2 - (\vec k \cdot \vec q)^2  \right)
\left(k^2 q^2-k_0 q_0 kq \right)^2}
{\vec k^2 \vec q^2 \vec p^2 k^2 p^2} \\
N_2(k,q)&= -2\frac{ \left(\vec q^2 \vec k^2 -(\vec k \cdot \vec q)^2  \right)
\left(k^2 p^2-k_0 p_0 kp \right)^2}
{ \vec k^2 \vec q^2 \vec p^2 k^2 q^2} \\
N_L(k,q)&=-\frac{1}{2 \vec k^2 \vec q^2 \vec p^2 k^2 q^2 p^2}
\left[(\vec p \cdot \vec k)\vec q^2\left(k_0p^2+p_0k^2\right)-\right. \nonumber \\ 
&\left. (\vec p \cdot \vec q)\vec k^2\left(p_0q^2+q_0p^2\right)+ 
(\vec k \cdot \vec q)\vec p^2\left(k_0q^2-q_0k^2\right) \right]^2.\label{eq:gluonkernelslongitudinalend}
\end{align}

\section{Numerical method}\label{num_sec}

\subsection{The algorithm}\label{appB1}

The numerical method is based on approximating the unknown exact solution of the 
coupled system of integral equations by an iterative procedure. Fixed-point 
iteration works fine enough for simple cases, but in the present example suffers
from extremely poor convergence. We therefore employed the Newton-Raphson method 
\cite{Press:92}, which improves the convergence drastically, see {\it e.g.\/}
ref.~\cite{Atkinson:1997tu} (where this method has been employed for the 
first time for DSEs in the continuum) and ref.\ \cite{Maas:2005xh}.
In the symbolic notation of eq.~(\ref{sym-DSE}) we can write the DSEs as
\begin{align}
f(D_i) = \frac{1}{D_i} - \frac{1}{D(\mu^2)} - \Pi_D(k^2_i) + \Pi_D(\mu^2)= 0\,,
\end{align}
with $D_i = D(k_i)$.
The Newton-Raphson method then relies on linearly improved iteration steps 
\begin{align}
D_i^{\eta+1}         &=  D_i^{\eta} - \delta D_i^{\eta+1},  \qquad 
 \delta D_i^{\eta+1}  =  J^{-1}(D_i^{\eta}) f(D_i^{\eta}) \\
 \intertext{with}
 J(D_i^{\eta})&=\frac{\partial f(D_i^{\eta})}{\partial D_i^{\eta}}\; . 
\end{align} 
We employ the following convergence criterion:
\begin{align}
  \left |\frac{\delta D^{\eta+1}_i}{D^{\eta}_i} \right | < \varepsilon  \quad 
  {\rm for} \quad i \in 1\dots N \; .
\end{align}
Note that one should not use a convergence criterion for an averaged relative change in the variables 
as the dressing functions $D(k_i^2)$ are rather steep in the infrared, where only a few
points $k_i$ are evaluated. Thus one can have situations where the majority of 
variables $D_i$ with $k_i$ in the ultraviolet are converged, whereas those $D_i$ which
correspond to small $k_i$ still change drastically.

Furthermore, we observed that the algorithm converges much better when the Newton step 
is only performed partly when one is still far away from the exact solution. 
The weight $\omega_{\textrm{Newton}}$ is calculated as  
\begin{align}
  \omega_{\textrm{Newton}} = 0.99 \; \textrm{min}\left\{ \left |\frac{D^{\eta}_i}{\delta D^{\eta+1}_i} 
    \right |, \;\; i \in \{ 1, \dots N \} \right \} \; .
\end{align}
If $\omega_{\textrm{Newton}} > 1$ it is set to one. This condition also ensures that
the dressing functions are positive definite on all intermediate iteration steps.

We determine the inverse of the matrix $J(D_i^{\eta})$ approximately by 
employing a generalisation of 
the Gauss-Seidel method called 'successive overrelaxation' (SOR). This procedure is much
faster than the exact inversion of the matrix. An exact determination of 
$J(D_i^{\eta})^{-1}$ is not needed anyway, since the inverse determines just the next respective 
iteration step, but is not used in calculating the right hand side of the integral equations.  

The SOR-algorithm necessarily converges, if the matrix $J$ is strictly 
diagonally dominated, {\it i.e.\/}
\begin{align}
  \vert J_{ii} \vert \geq \epsilon + \sum^N_{\stackrel{j=1}{j\neq i}}\vert J_{ij} \vert \; .
\end{align}
The algorithm starts with an initial guess for the solution of the linear equation 
$Jx=b$ with $x,b \in \mathbb{R}^N$ and $J \in \mathbb{R}^{N\times N}$. This guess
is improved at each SOR-step $\eta \rightarrow \eta+1$ according to:
\begin{align}
  x^{\eta+1}_i=(1-\omega)x^{\eta}_i+ \frac{\omega}{J_{ii}}
  \left( b_i - \sum^{i-1}_{j=1} J_{ij} x^{\eta+1}_j 
  -\sum^N_{j=i+1} a_{ij} x^{\eta}_j  \right).
\end{align}
The relaxation parameter $\omega$ interpolates linearly between the old vector
$x^{\eta}$ and the Gauss-Seidel rule $\omega/J_{ii} (\cdots)$. The algorithm converges 
for values of
$\omega$ in the range $0<\omega <2$. If the matrix $J$ is strictly diagonally 
dominant, then convergence is best for $1<\omega<2$ ({\it i.e.} 'overrelaxation').
However, other cases may give best results for a value between $0<\omega<1$ 
('underrelaxation'). We tried values for $\omega$ between $0.7$ and $1.2$ and obtained
best results for $w\approx 0.8$. This clearly indicates, that the Jacobian matrix is not 
strictly diagonally dominant but still well behaved enough
for the SOR algorithm to converge.
For the update of the i-th component of $x$, its relative change is  
\begin{align}
\delta_i^{\eta+1} = \frac{\omega}{b_i}\left( b_i - \sum^{i-1}_{j=1} J_{ij} x^{\eta+1}_j -
     \sum^N_{j=i} J_{ij} x^{\eta}_j  \right) \; . 
\end{align}    
These quantities are used to calculate an estimated averaged error from the sum
of the absolute values of the relative changes in the current approximation to the solution
\begin{align}
 \varepsilon^{\eta+1} = \frac{1}{N}\sum^N_{i=1} \left | \delta^{\eta+1}_i \right | \; .      
\end{align}
The criterion for convergence is, that $\varepsilon^{\eta+1} < \epsilon$,
for a given small $\epsilon$.

In our implementation of the Newton-Raphson method the SOR-algorithm has to be called
at every step of the Newton-iteration. In general it delivers a converged result for 
$J^{-1}$. In the rare occasions where the SOR-algorithm did not converge, we determined 
$J^{-1}$  with a Gauss-Jordan elimination and returned to the SOR-algorithm in the
next Newton-step.

\subsection{Start guess for the dressing functions}\label{sec:startguess}

The Newton-Raphson iteration method described in the last section is a local method, which
converges quadratically, if the start guess for the dressing functions is close enough to
the final solutions. The exact meaning of 'close enough', however, depends very much on 
the problem at hand. If the dressing functions are infrared finite, as for example in the
Dyson-Schwinger equations for the quark propagator \cite{Maris:2003vk,Fischer:2003rp}, then
convergence is achieved for almost any form of the start guess. However, if the dressing
functions are infrared singular, as in the present case, much more care is required and
one should provide a start guess that is qualitatively similar to the actual 
solution\footnote{As an alternative one can resort to a globally converging method, see
ref.~\cite{Maas:2005xh}.}. In our case such a start guess can be provided by noting
that the continuum version of the DSEs (\ref{eq:ghostlong}),(\ref{eq:gluontransverse}) 
and (\ref{eq:gluonlongitudinal}) can be solved analytically in the zero temperature
limit for both very large and very small momenta. 

In the infrared asymptotic limit, the ghost and gluon dressing functions, $G(p^2)$ and $Z(p^2)$, are 
proportional to simple power laws \cite{vonSmekal:1997is}
\beqa
Z(p^2) \sim  (p^2)^{2\kappa}, \qquad 
G(p^2) \sim  (p^2)^{-\kappa}, \label{powerlaw} 
\eeqa
with $\kappa = \frac{93-\sqrt{1201}}{98} \approx 0.595353$ \cite{Zwanziger:2001kw,Lerche:2002ep}. 
Correspondingly the running coupling (\ref{alpha}) has an infrared fixed point, 
\beq
\alpha(0) = \frac{4 \pi}{6N_c}
\frac{\Gamma(3-2\kappa)\Gamma(3+\kappa)\Gamma(1+\kappa)}{\Gamma^2(2-\kappa)
\Gamma(2\kappa)} \approx \frac{8.92}{N_c}. \label{fixedpoint}
\eeq
Both, the values of the exponent $\kappa$ and the fixed point $\alpha(0)$ have recently
been shown to be independent of the gauge parameter in a class of gauges that interpolate
between Landau and Coulomb gauge \cite{Fischer:2005qe}. 

In the ultraviolet asymptotic limit one recovers the well known expressions from resummed 
perturbation theory,
\begin{eqnarray}
Z(p^2) &=& Z(\mu^2) \left[ \frac{11 N_c \alpha(\mu^2)}{12 \pi} \log\left(\frac{p^2}{\mu^2}\right)+1 \right]^\gamma  \; ,
\label{gluon_uv}\\
G(p^2) &=& G(\mu^2) \left[ \frac{11 N_c \alpha(\mu^2)}{12 \pi} \log\left(\frac{p^2}{\mu^2}\right)+1 \right]^\delta  \; .
\label{ghost_uv}
\end{eqnarray}
Here $Z(\mu^2)$ and $G(\mu^2)$ denote the value of the dressing functions at some
renormalisation point $\mu^2$ and $\gamma$ and $\delta$ are the respective 
anomalous dimensions. To one loop order one has $\delta = - 9/44$ and $\gamma = - 1
-2\delta=-13/22$ for arbitrary number of colours $N_c$ and no quarks, $N_f=0$. 

\begin{table} \label{tab:parms}
\begin{center}
$\begin{array}{|c|c|c|c|c|c|}
\hline
\textrm{Parameters} &    C     &    D_s   &    \beta  &  \nu   & \eta   \\ \hline
G(p^2)              & \ 0.35 \ & \ 0.02 \ &\   -0.3\  &\ 0.5\  & \ -0.2 \   \\
Z_m(p^2),Z_0(p^2)   &    550   &   0.01   &     1.2   &  2.0   &   -0.6 \\\hline
\end{array}
$
\caption{Parameters for the start guesses for the gluon and ghost dressing functions.}
\end{center}
\end{table}

A simple ansatz that interpolates between both limits is given by 
\begin{align}
F(p^2) = C \left(\frac{p^2}{p^2/D_s+1} \right)^{\beta} \biggl( \nu \log\left(p^2/D_s+1\right)+1 \biggl)^{\eta} \; ,
\end{align}
where $C,D_s,\beta,\nu, \eta$ are parameters to be chosen appropriately. Our choice of parameters for the
ghost dressing function and the heat bath transverse and longitudinal dressing of the gluon 
are listed in table 3. As long as the qualitative properties of the start guess remains unchanged
for the relevant momenta convergence is achieved for a whole range of parameter choices centred
around the values given.

\end{appendix}

\end{document}